\documentclass[conference]{IEEEtran}

\IEEEoverridecommandlockouts

\usepackage{cite}
\usepackage{amsmath,amssymb,amsfonts}
\usepackage{algorithmic}
\usepackage{graphicx}
\usepackage{textcomp}

\usepackage{multirow}
\usepackage{booktabs}
\usepackage{float}
\usepackage[
  colorlinks=true, linkcolor=blue, urlcolor=blue, citecolor=red, pdfborder={0 0 0}
]{hyperref}

\newcommand{\cmmnt}[1]{\ignorespaces}

\def\BibTeX{{\rm B\kern-.05em{\sc i\kern-.025em b}\kern-.08em
    T\kern-.1667em\lower.7ex\hbox{E}\kern-.125emX}}

\begin{document}
\title{Unveiling Urban Mobility Patterns: A Data-Driven Analysis of Public Transit \\
  \thanks{This research was funded by the PERSEUS project under the Marie Skłodowska-Curie grant agreement No. 101034240.}
}
\author{
  \IEEEauthorblockN{Oluwaleke Yusuf, Adil Rasheed}
  \IEEEauthorblockA{Department of Engineering Cybernetics\\
    Norwegian University of Science\\and Technology\\
    7034 Trondheim, Norway\\
    Email: oluwaleke.u.yusuf@ntnu.no, adil.rasheed@ntnu.no}
  \and
  \IEEEauthorblockN{Frank Lindseth}
  \IEEEauthorblockA{Department of Computer Science\\
    Norwegian University of Science\\and Technology\\
    7034 Trondheim, Norway\\
    Email: frankl@ntnu.no}
  \and
  \IEEEauthorblockN{Martin Slaastuen}
  \IEEEauthorblockA{Market Insight\\
    AtB AS\\
    7011 Trondheim, Norway\\
    Email: martin.slaastuen@atb.no}
}
\maketitle

\begin{abstract}
  The expansion of urban centers necessitates enhanced efficiency and sustainability in their transportation infrastructure and mobility systems. The big data obtainable from various transportation modes potentially offers critical insights for urban planning.
  This study presents analysis of detailed historical public transit data, enriched with relevant temporal and geospatial metadata, as a precursor to injecting dynamism into digital twins of mobility systems via ML/DL-based predictive modeling.
  A data preprocessing framework was implemented to refine the raw data for effective historical analysis and predictive modeling. This paper examines public transit data for patterns and trends---incorporating factors such as time, geospatial elements, external influences, and operational aspects.
  From a technical standpoint, this research helps to assess the quality of the available transit data and identify important information for use in digital twins. Such digital twins foster educated decisions for efficient, sustainable urban mobility systems by anticipating infrastructure demand, identifying service gaps, and understanding mobility dynamics.
\end{abstract}


\section{Introduction}
Trondheim, the third most populous municipality in Norway, is an ever-growing urban center in Trøndelag county, home to approximately 206,000 inhabitants as of 2023. This growth---fueled by its status as an educational and research hub with institutions like the Norwegian University of Science and Technology (NTNU), SINTEF, and St. Olav's University Hospital---places increasing demands on its transportation infrastructure. With the resultant increase in car traffic and air pollution, the focus of city planners has shifted towards sustainable transport, with initiatives like Miljøpakken and MobilitetsLab Stor-Trondheim emphasizing public transport and active mobility over driving.

The increasing reliance on public transit in urban areas like Trondheim has spurred significant academic interest in analyzing public transit data. Researchers utilize varied data sources --- including Automated Passenger Counting (APC) \cite{Liu2021mro}, General Transit Feed Specification (GTFS) \cite{Zhang2020ipp, Liu2021mro, Liu2023rae, Park2020apt}, Integrated Circuit (IC) card \cite{Huang2023rao, Li2020ptb, Cottreau2023stp} --- to glean insights into urban transit efficiency, punctuality, and service reliability. These analyses inform infrastructure planning and policy-making, contributing to the development of more efficient and sustainable transportation infrastructure and mobility systems.

Recent literature has explored diverse aspects of public transit. Studies like Zhang et al. \cite{Zhang2020ipp} have identified key transit corridors using smart card data, while Yao et al. \cite{Yao2019etp} introduced models to measure public transit systems' performance, considering multiple stakeholders and environmental factors. Huang et al. \cite{Huang2023rao} proposed frameworks for analyzing bus network resilience, and Li et al. \cite{Li2020ptb} examined passenger travel behavior in urban transit corridors. These studies, along with others \cite{Liu2021mro, Liu2023rae, Park2020apt, Cottreau2023stp}, provide a foundation for understanding the complexities of urban mobility and the need for data-driven approaches for infrastructure planning and policy-making to enhance public transit systems.

Within this context, a Digital Twin (DT) provides a comprehensive technological framework for aggregating real-time data from transportation infrastructure and mobility systems into a unified digital representation. By integrating real-world information with a visual representation of the entire system, the DT achieves a realistic mimicry of the system's state and behavior. This level of digitalization allows for in-depth analysis of historical trends for decision-making, real-time analytics for automated responses and emergency coordination, predictive modelling for forecasting and advanced simulations for exploratory "What-if?" analyses.

However, the intricate spatiotemporal dynamics of urban mobility systems and the interplay between various modes of transportation make analyzing past data and forecasting future trends a daunting task. As a foundation for traffic modeling similar to \cite{Saki2020ana} for applications in a DT of Trondheim's mobility system, this study seeks to tackle these challenges by initially focusing on the historical analysis of public transit data. This analysis assesses the utility of such data in understanding the system's dynamics and interdependencies to facilitate its automation through a DT.

\section{Data}
This study is based on data obtained from AtB, Trøndelag's public transport administrator, via APC systems installed on buses---which employ DILAX optical sensors with 3D stereoscopic vision technology---for passenger counting. Despite occasional technical issues leading to counting discrepancies \cite{Liabo2023mpc}, the APC data, after preprocessing, reliably reflects passenger trends and patterns across the bus transit system.

\begin{figure}[t]
  \centering\includegraphics[width=0.99\linewidth]{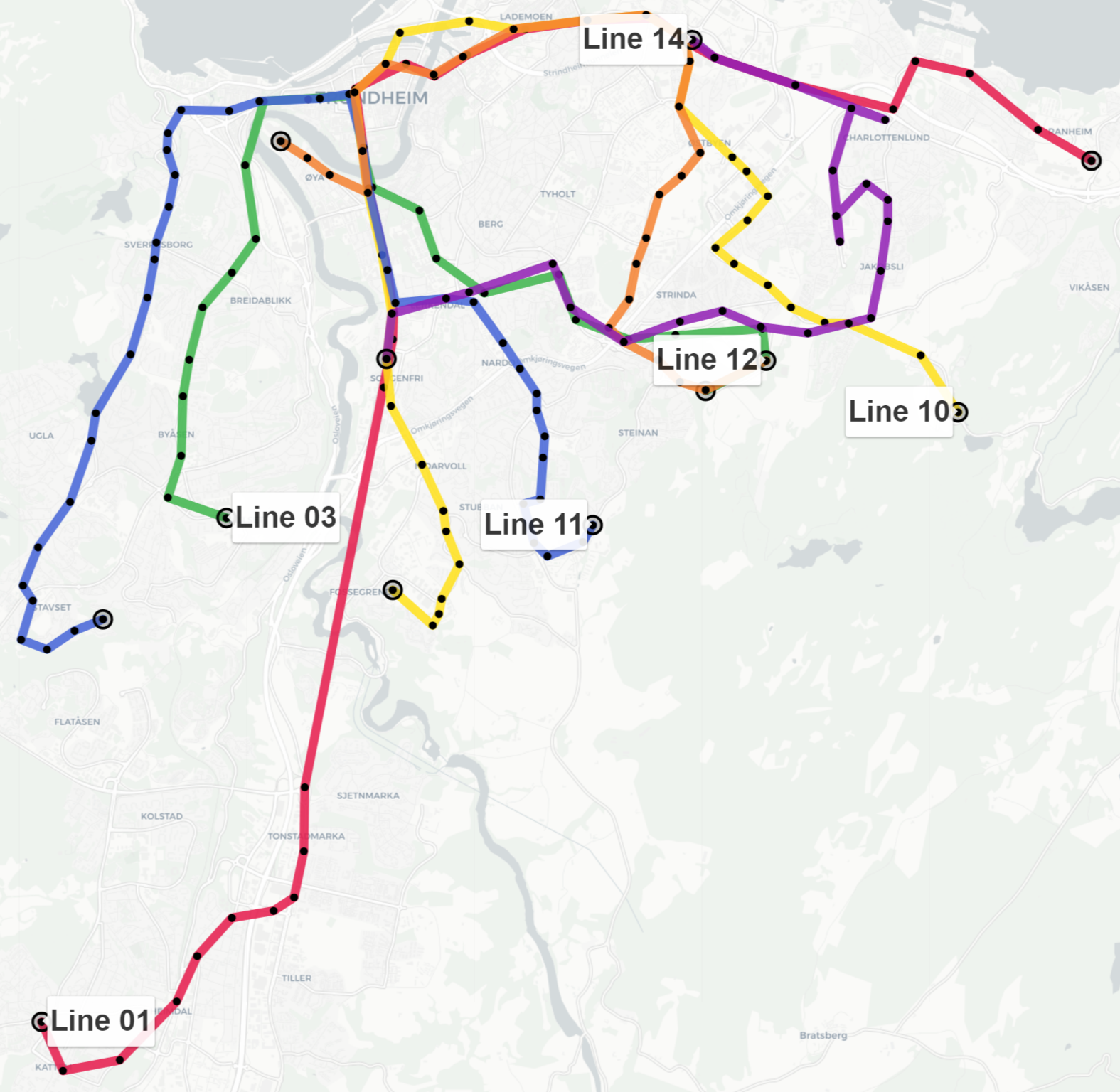}
  \vspace{-15pt}\caption{Satellite map of the AtB bus lines in the dataset.}
  \vspace{-10pt}\label{fig:map-all-lines}
\end{figure}

\subsection{Data Description}
The study's dataset covers the period from 1st May 2020 to 30th November 2023, comprising 1,179,770 unique trips over 1,295 days, across 6 lines and 204 stops in Trondheim Municipality. As shown in \autoref{fig:map-all-lines}, the bus lines in the data cover a wide section of the city, with most of them routing through the city center. The raw dataset features a comprehensive set of 63 attributes, offering insights into the temporal and spatial aspects of bus trips, passenger flow metrics, and various operational parameters.

\begin{itemize}
  \item \textbf{Temporal Information}: Includes date and time features, such as \textit{TripScheduledDeparture}, \textit{TripDepartureHour}, \textit{StopScheduledArrival} and \textit{StopActualArrival}. These capture the scheduled and actual timings for departures and arrivals for specific stops and entire trips.
  \item \textbf{Operational Flags}: A series of conditional flags such as \textit{StopTime}, \textit{FLAG\_Trip\_15minDeviation} and \textit{FLAG\_Stop\_IsDelayed} offer insights into operational issues.
  \item \textbf{Spatial Information}: Encoded via \textit{StopName}, \textit{Longitude}, and \textit{Latitude}, this provides a spatial context to each trip.
  \item \textbf{Passenger Flow Metrics}: The \textit{Boarding}, \textit{Alighting}, and \textit{Onboard} fields represent data from the APC system, which records the volume of passenger traffic.
  \item \textbf{Vehicle and Service Information}: Details such as \textit{BusType} and \textit{CarriageNumber} provide data on the assets used and the .
  \item \textbf{Special Dates}: The dataset is enriched with flags such as \textit{FLAG\_Holiday\_Restday} and \textit{FLAG\_SchoolVacation} and descriptive fields such as \textit{DayComment} and \textit{HolidayName} to denote public holidays and school breaks, allowing for nuanced analyses of service variations on special days.
\end{itemize}

\subsection{Data Preprocessing}
As previously mentioned, the dataset's foundation is the APC data, enriched with pertinent operational and spatiotemporal details. The raw dataset was preprocessed to address missing values across the dataset and correct negative values in the APC and temporal data, as outlined below:

\begin{figure*}[h]
  \centering\includegraphics[width=0.99\linewidth]{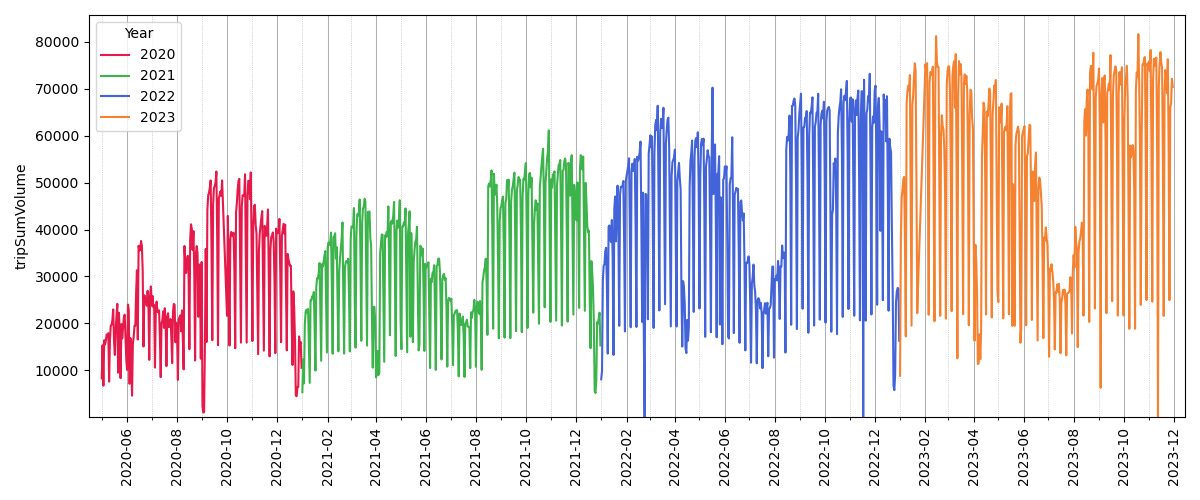}
  \vspace{-10pt}\caption{Yearly breakdown of passenger volume for the entire dataset from May 2020 to November 2023.}
  \vspace{-10pt}\label{fig:Date-tripSumVolume-Year}
\end{figure*}

\begin{figure*}[h]
  \centering\includegraphics[width=0.99\linewidth]{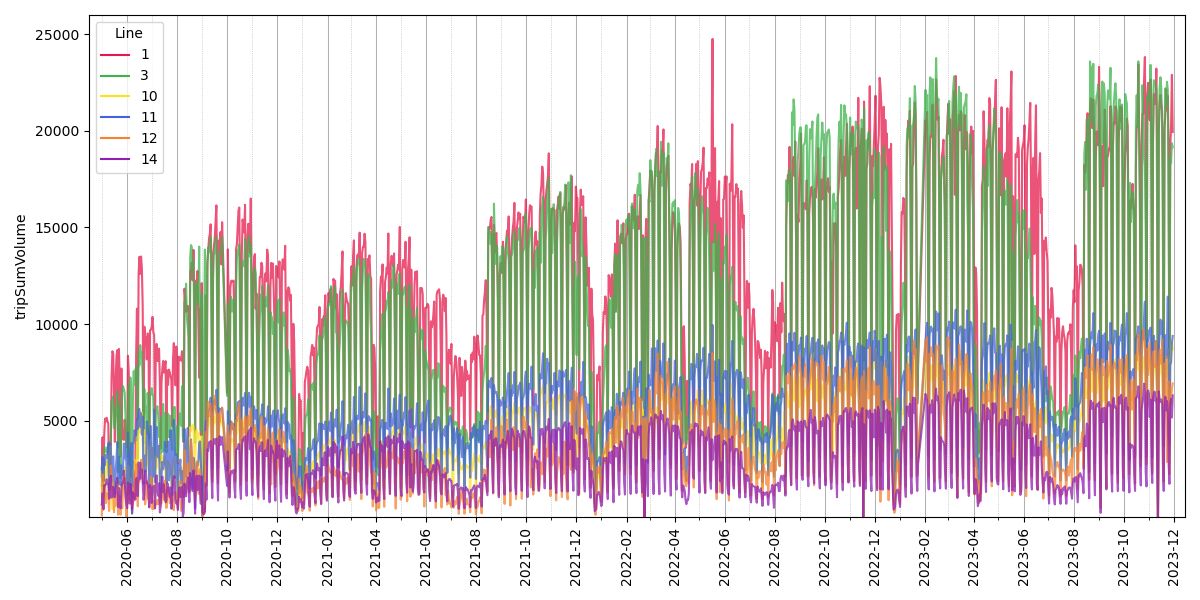}
  \vspace{-10pt}\caption{Detailed overview of passenger volume across different bus lines for the entire dataset.}
  \vspace{-10pt}\label{fig:Date-tripSumVolume-Line}
\end{figure*}

\begin{enumerate}
  \item \textit{APC Features}: The dataset primarily includes the \textit{Boarding} and \textit{Alighting} features from the APC system, representing passengers getting on and off the bus at each stop. These figures are aggregated from counts at each door. Derived from these, new metrics like passenger volume on the bus between stops and the change in passenger count throughout the trip are calculated.

        Data cleaning involved: \textit{(i)} Setting missing values in \textit{Boarding} and \textit{Alighting} to zero. \textit{(ii)} Discarding erroneous derivative APC values and engineering three new features \textit{busVolume}, \textit{tripSumVolume}, and \textit{stopVolume} which denote the passenger count onboard, cumulative boardings since the trip's start, and count at each stop, respectively. \textit{(iii)} Trips with negative \textit{Boarding} or \textit{Alighting} values or where flags \textit{FLAG\_Trip\_APCActive} or \textit{FLAG\_Trip\_APCExtremeValue} were \textit{True} were excluded from the dataset.

  \item \textit{Temporal Information}: For bus schedule features \textit{TripScheduledArrival}, \textit{StopActualArrival}, and \textit{StopActualDeparture}, empty values were replaced with corresponding scheduled values. Instances where \textit{StopActualDeparture} was earlier than \textit{StopActualArrival} were corrected and the \textit{StopTime} subsequently recalculated to reflect this correction.

  \item \textit{Missing Values}: Missing values for the conditional operational flags were determined using expert knowledge and set accordingly. Missing \textit{BusType}, \textit{DayComment}, and \textit{HolidayName} values were filled with ``No Information", while modal values were used for missing \textit{Municipality} and \textit{StopType} data.

  \item \textit{Unique Trip Identifiers}: The \textit{Trip} feature, a daily unique identifier, was combined with the \textit{Date} feature to create a dataset-wide unique identifier \textit{dateTrip}.
\end{enumerate}

In the final preprocessing step, we removed features deemed irrelevant, outdated (as confirmed by expert consultation), or redundant. The resultant dataset includes 45 features and 1,130,100 \cmmnt{1,112,221} unique trips, representing a 4.21\% reduction from the original dataset. This reduction primarily addresses inaccuracies in the APC data, notably from May to September 2020, a period shortly after AtB deployed the APC system, and its bugs were being ironed out.

\section{Analytics and Discussion}
This section interprets the intricate dynamics of Trondheim's bus transit system through APC data, focusing on passenger volume over time. The analysis encompasses temporal fluctuations, the impact of external factors, and operational efficiencies within the network.

\subsection{Temporal Patterns}
\label{sec:temporal-patterns}
The temporal analysis of the data is performed at various levels---hourly, daily, weekly, and monthly---using the mean value for each level. This analysis specifically considers data from 1st March 2022 to 30th November 2023, excluding the Covid-19 period, which is examined separately in \autoref{sec:external-influences}.

\begin{figure*}[h]
  \begin{minipage}{\columnwidth}
    \centering\includegraphics[width=0.99\linewidth]{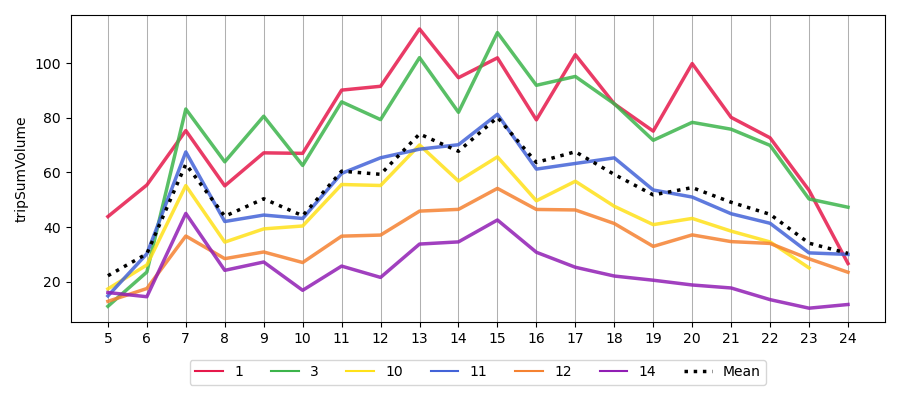}
    \vspace{-20pt}\caption{Hourly variation of passenger volume across bus lines.}
    \vspace{-10pt}\label{fig:TripDepartureHour-tripSumVolume-Line}
  \end{minipage}%
  \hfill 
  \begin{minipage}{\columnwidth}
    \centering\includegraphics[width=0.99\linewidth]{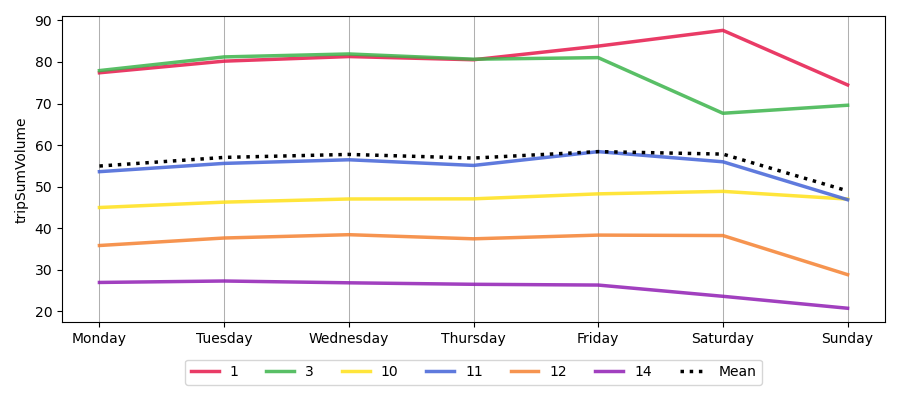}
    \vspace{-20pt}\caption{Daily variation of passenger volume across bus lines.}
    \vspace{-10pt}\label{fig:DayType-tripSumVolume-Line}
  \end{minipage}
\end{figure*}

\begin{figure*}[h]
  \begin{minipage}{\columnwidth}
    \centering\includegraphics[width=0.99\linewidth]{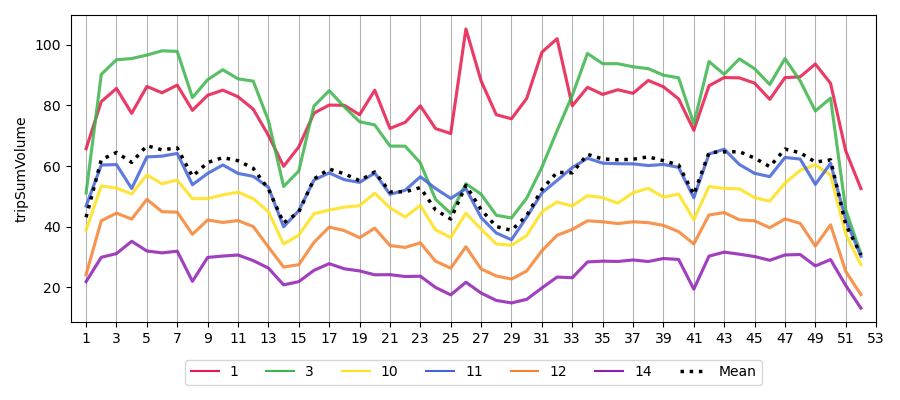}
    \vspace{-20pt}\caption{Weekly variation of passenger volume across bus lines.}
    \vspace{-10pt}\label{fig:WeekNumber-tripSumVolume-Line}
  \end{minipage}%
  \hfill 
  \begin{minipage}{\columnwidth}
    \centering\includegraphics[width=0.99\linewidth]{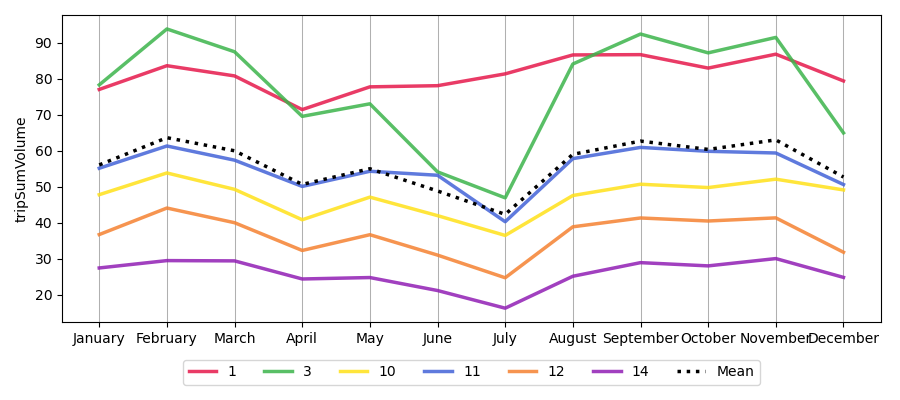}
    \vspace{-20pt}\caption{Monthly variation of passenger volume across bus lines.}
    \vspace{-10pt}\label{fig:Month-tripSumVolume-Line}
  \end{minipage}
\end{figure*}

\subsubsection{Total Daily Volume}
\label{sec:total-daily-volume}
Analyzing the total daily passenger volume for the entire system from 1st May 2020 until 30th November 2023, as illustrated in \autoref{fig:Date-tripSumVolume-Year}, reveals a general upward trend in passenger volume, influenced by Covid-19 recovery and population increases in Trondheim. The data shows notable dips in volume during holidays such as Easter, Summer, Christmas, and New Year. Additionally, the highest daily volumes correspond to specific events like the annual UKA Student Festival and St. Valentine's Day.

Further breaking down the daily passenger volume by line, as shown in \autoref{fig:Date-tripSumVolume-Line}, highlights variations in the passenger volume experienced by each bus line over time. Line 1 handled a total volume of 16,050,047 passengers, closely followed by Line 3 at 14,282,754 passengers with Line 11 a distant third at 7,169,532 passengers. Lines 1 and 3 exhibit distinct patterns, particularly around major holiday periods, with Line 3’s volume affected by the NTNU academic calendar.

\subsubsection{Average Hourly and Daily Volumes}
The average hourly passenger volumes, displayed in \autoref{fig:TripDepartureHour-tripSumVolume-Line}, show a consistent morning peak at 07:00 across all lines, with an additional surge in Line 3 due to NTNU traffic. The afternoon peaks are dispersed between 14:00 and 16:00, and a third peak at 20:00 is more pronounced for Line 1, likely due to its proximity to dining and entertainment areas.

The daily passenger volumes, as depicted in \autoref{fig:DayType-tripSumVolume-Line}, are relatively stable throughout the week with a noticeable dip on Sundays. Lines 1 and 3 display contrasting weekend patterns, possibly reflecting a shift in student traffic on weekends.

\subsubsection{Average Weekly and Monthly Volumes}
The average weekly and monthly passenger volumes, as detailed in Figures \ref{fig:WeekNumber-tripSumVolume-Line} and \ref{fig:Month-tripSumVolume-Line}, generally mirror each other, with weekly data offering more granular insights. Lines 10, 11, 12, and 14 exhibit similar trends at both the weekly and monthly levels. Seasonal trends are evident, such as dips during holidays and spikes in travel activity at specific times of the year. Lines 1 and 3 show deviations from these patterns due to their unique passenger demographics and the seasonal impact of NTNU’s academic schedule.

\subsection{External Influences}
\label{sec:external-influences}
This subsection delves into how external events, including the COVID-19 pandemic and major holidays, influence the daily passenger volume across Trondheim’s bus transit system.

\subsubsection{Covid-19 Pandemic}
\begin{figure*}[tb]
  \centering\includegraphics[width=0.99\linewidth]{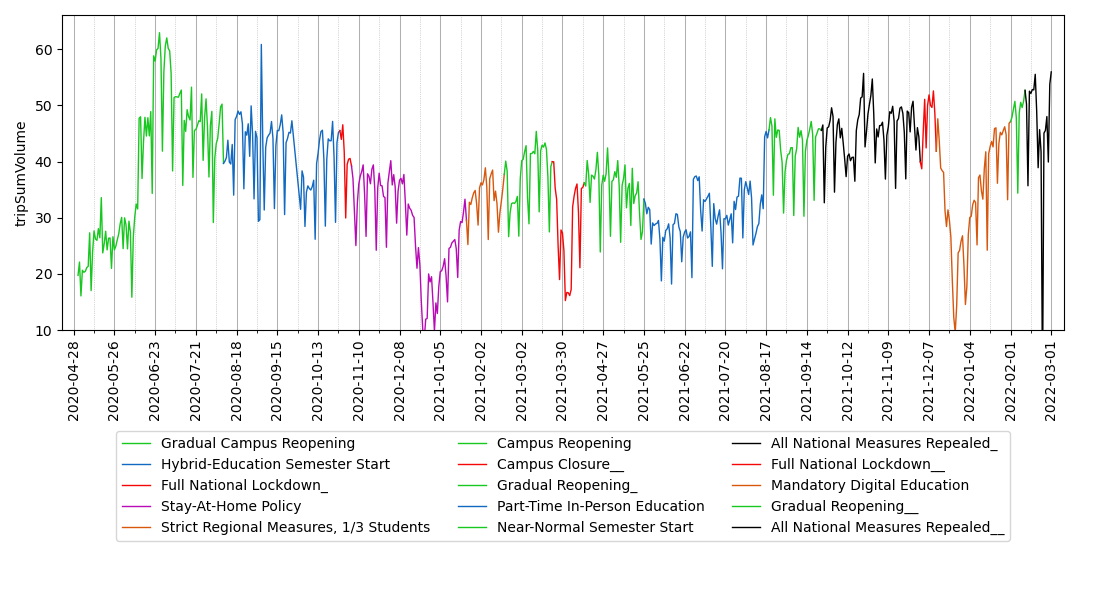}
  \vspace{-30pt}\caption{Effect of Norwegian Covid-19 policies on passenger volume from May 2020 to March 2022.}
  \vspace{-10pt}\label{fig:Date-tripSumVolume-CovidPeriod}
\end{figure*}

The pandemic's impact on passenger volumes is evident in Figures \ref{fig:Date-tripSumVolume-Year} and \ref{fig:Date-tripSumVolume-Line}. The Norwegian government's Covid-19 policies, when mapped against the bus transit data as displayed in \autoref{fig:Date-tripSumVolume-CovidPeriod}, show a clear correlation between policy changes and passenger volumes, justifying the exclusion of this period from earlier analyses.

\subsubsection{Major Holidays and Local Events}
\begin{figure*}[tb]
  \centering\includegraphics[width=0.99\linewidth]{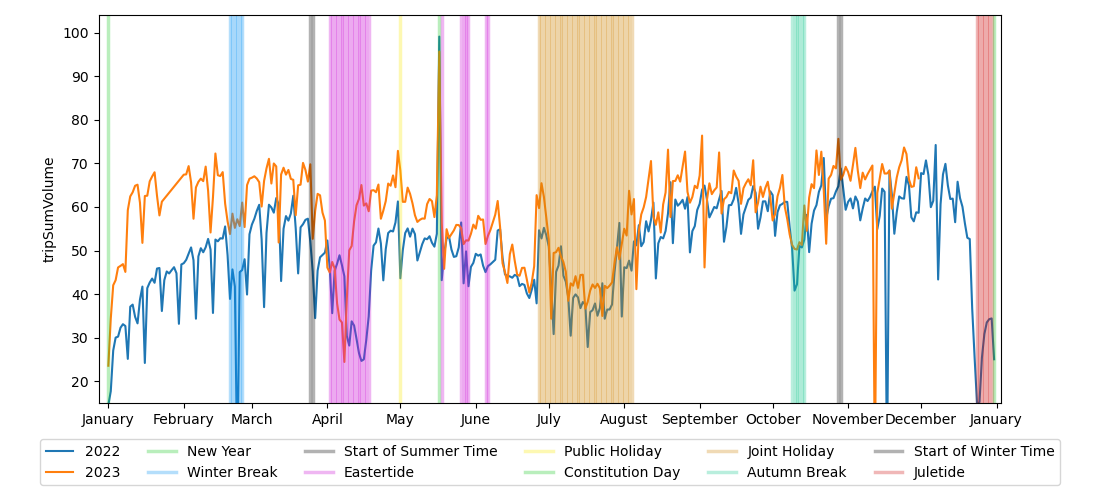}
  \vspace{-10pt}\caption{Variation of passenger volume with major holidays and local events in 2022 and 2023.}
  \vspace{-10pt}\label{fig:Date-tripSumVolume-DayComment}
\end{figure*}

To understand the effects of holidays and local events on average daily passenger volumes, the data for 2022 and 2023 is plotted with overlaid colored bands marking significant dates, as seen in \autoref{fig:Date-tripSumVolume-DayComment}. Generally, passenger volume drops on these dates, with the notable exception of ``Constitution Day," which causes a surge in volume. The comparison between 2022 and 2023 shows initial discrepancies, likely due to COVID-19's lingering effects, but a gradual alignment in patterns as the year progresses.

\subsection{Operational Trends}
\subsubsection{Trip Counts and Passenger Volumes}
\label{sec:trip-counts-passenger-volumes}
\begin{figure*}[t]
  \begin{minipage}{\columnwidth}
    \centering\includegraphics[width=0.99\linewidth]{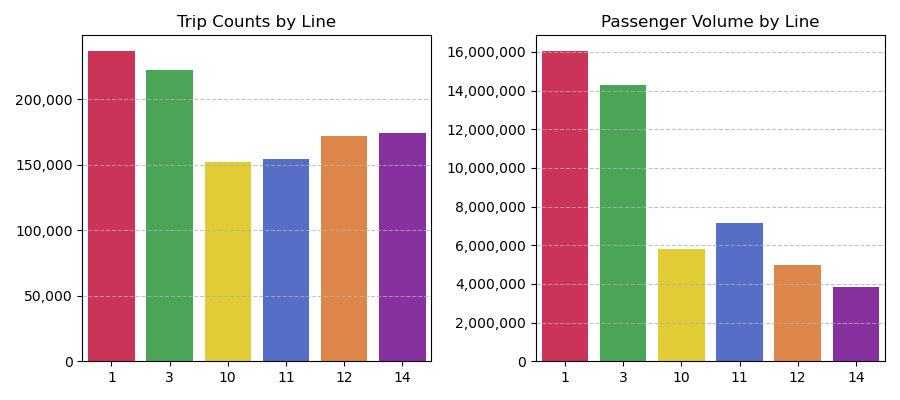}
    \vspace{-20pt}\caption{Comparison of trip counts and passenger volumes handled by different bus lines.}
    \vspace{-10pt}\label{fig:Line-dateTrip-tripSumVolume}
  \end{minipage}%
  \hfill 
  \begin{minipage}{\columnwidth}
    \centering\includegraphics[width=0.99\linewidth]{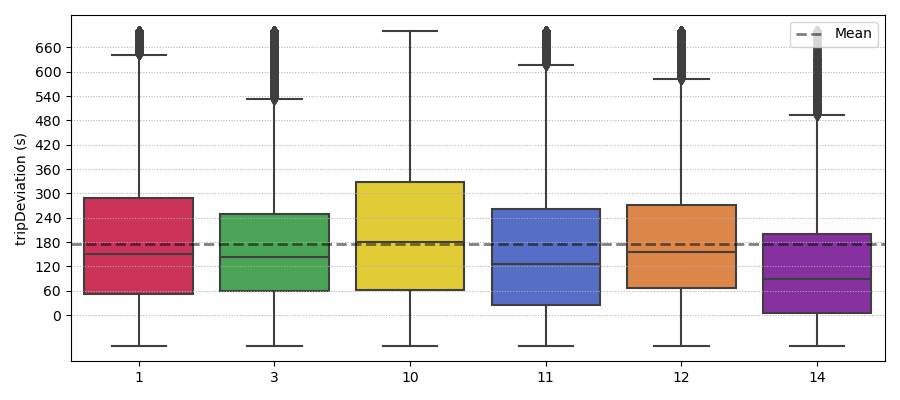}
    \vspace{-20pt}\caption{Trip scheduling deviations for different bus lines.}
    \vspace{-10pt}\label{fig:Line-tripDelay}
  \end{minipage}
\end{figure*}

\autoref{fig:Line-dateTrip-tripSumVolume} presents the total bus trips and passenger volumes for each line. Echoing findings from \autoref{sec:total-daily-volume}, Line 1 manages the highest number of trips and passengers, closely followed by Line 3. A noteworthy observation is the contrast between the trip counts and passenger volumes for Lines 10, 11, 12, and 14. Lines 11 and 10, despite similar trip counts, see Line 11 carrying more passengers. A similar disparity exists between Lines 12 and 14, with Line 14 transporting more passengers. This indicates a potential misalignment between the scheduling supply and actual passenger demand on Lines 10 and 14, often resulting in underutilized buses.

\autoref{tab:BusType-dateTrip-tripSumVolume} ranks the top 5 bus types used in Trondheim by AtB, based on passenger volume and trip count. Dominating this list are the Van Hool ExquiCity buses, primarily servicing the high-demand Lines 1 and 3. This table also sheds light on AtB's efforts to enhance the sustainability of the bus transit system with hybrid, electric, and biofuel buses, though fossil fuels remain predominant. This reliance is an area of concern for future adaptation to the growing population and increasing demand placed on the bus system.

Further analysis of individual vehicles in the AtB fleet shows the top-10 chassis are all Van Hool ExquiCity 24m Diesel-Hybrid models, predominantly operated on Line 3. These buses, averaging 15,925 trips and conveying 1,035,355 passengers, are the fleet's workhorses. Their significant usage underscores the importance of rigorous maintenance schedules to extend their operational lifespan and ensure system reliability.

\begin{table}[h]
  \caption{Comparison of trip counts and passenger volumes handled by different bus types.}
  \label{tab:BusType-dateTrip-tripSumVolume}
  \begin{tabular}{lllrr}
    \toprule
    Bus Type              & Fuel          & Trips   & Volume     \\
    \midrule
    Van Hool ExquiCity    & Diesel-Hybrid & 457,964 & 30,250,790 \\
    MAN Lion's City 18    & Diesel        & 334,790 & 11,381,406 \\
    MAN Lion's City G A23 & Biodiesel/CNG & 129,001 & 5,023,050  \\
    Heuliez GX 437        & Electric      & 154,053 & 4,440,077  \\
    MAN Lion's City A21   & Biogas/CNG    & 22,241  & 746,516    \\
    \bottomrule
  \end{tabular}
  \vspace{-15pt}
\end{table}

\subsubsection{Trip Deviations and System Efficiency}
\label{sec:trip-deviations-system-efficiency}
A key aspect of assessing the bus transit system's efficiency is examining the deviations between scheduled trip completion times and actual end times, i.e., the bus's arrival at the last stop. These trip-level deviations are an aggregate of cumulative stop-level deviations along the route. Positive deviations indicate late arrivals, while negative ones signify early arrivals. \autoref{fig:Line-tripDelay} illustrates the distribution of these trip deviations in seconds for each line. The system-wide average deviation stands at 176 seconds, with Line 14 performing best at an average of 120 seconds, and Line 10 lagging with an average deviation of 217 seconds.

The root causes of trip deviations can vary, with passenger volume and road traffic being prime factors. Line 14, which has the least passenger volume, consequently spends less time at stops, leading to its lower deviation. However, an interesting anomaly arises with Line 10. As discussed in \autoref{sec:trip-counts-passenger-volumes}, despite having a lower passenger volume compared to Line 11, Line 10's average deviation is significantly higher. This discrepancy suggests that other factors, possibly specific to Line 10's route or operational dynamics, contribute to its higher deviation, warranting further investigation to fully understand these patterns.

\section{Conclusion and Future Work}
\subsection{Conclusion}
This paper analyzed mobility data from Trondheim's public bus transportation system spanning from May 2020 until November 2023. It sheds light on the shifting dynamics of passenger volume against the backdrop of the city's growing population and escalating demands on the bus system.
The study of temporal patterns across specific bus lines offers insights into the characteristics of various city areas and their demographics, particularly as said areas and demographics evolve over time. In addition, our analysis reveals the impact of external factors, notably the varied government responses to the COVID-19 pandemic, on mobility trends.
This research contributes to a broader research project that aims to create a multi-modal Digital Twin (DT) of Trondheim’s mobility system, instilled with dynamism via real-world data, for enhancing the efficiency and sustainability of future urban mobility solutions.

\subsection{Future Work}
Future work aims at developing automated AI-based solutions for mobility systems, incorporating various machine (and deep) learning techniques for geospatial analysis, dimensionality reduction, predictive modelling and synthetic data generation for use in DTs. This initiative will involve the use of multi-modal mobility data, thereby augmenting the DT with extensive data-driven historical analyses and predictive models across the entire transportation infrastructure and mobility system.
Additionally, enhancements will be made to the accuracy of the APC data by integrating onboard camera footage for more precise passenger counts. This enhanced data will underpin trip-level and stop-level predictive models for passenger volume at various temporal scales. The resulting insights and models will be instrumental in strengthening the resilience, sustainability, and efficiency of Trondheim's bus transit system as it adapts to an increasing population.

\section*{Acknowledgment}
This research received funding from the PERSEUS Doctoral Program, supported under the Marie Skłodowska-Curie grant agreement No. 101034240. The authors also acknowledge MobilitetsLab Stor-Trondheim for their financial contribution. We are grateful to AtB for providing the essential mobility data for our research and extend special thanks to Mats Lien at AtB for his assistance.

\bibliographystyle{AR}
\bibliography{bibliography}

\end{document}